# Extension of interferometric particle imaging to small ice-crystal sizes using the Discrete Dipole Approximation

Marc Brunel[1], Gilles Demange[2], Renaud Patte[2], Maxim Yurkin[1]

1 : Rouen Normandie Université, INSA de Rouen, CNRS, UMR 6614 CORIA, Avenue de l'Université, 76801 Saint-Etienne du Rouvray cedex, France

2 : Rouen Normandie Université, INSA de Rouen, CNRS, UMR 6634 GPM, Avenue de l'Université, 76801 Saint-Etienne du Rouvray cedex, France

Abstract

Interferometric Particle Imaging (IPI) is a powerful technique to characterize aerosol particles, which so far has been applied only to particles larger than about 100 wavelengths. We extend its applicability to smaller ice crystals, combining rigorous modelling of particle shapes with the Phase Field Modelling and light-scattering simulations with the Discrete Dipole Approximation (DDA). Even for particles with the largest dimension of 11.5 wavelengths (and the smallest one comparable to the wavelength), the 2D Fourier transform of the interferometric image remains linked to the 2D autocorrelation of the particle shape at various viewing angles, validating the general measurement principle. However, the sensor must necessarily have wide viewing angle, which complicates interpretation of apparent particle shape, when such particles are observed from the edge. IPI is, thus, shown to be a powerful optical technology for characterizing ice particles down to a few micrometers in the atmosphere. Meanwhile, DDA is a versatile method for such synthetic experiments and can further supply large datasets for development of various inversion methods.

Introduction

Interferometric Laser Imaging for Droplet Sizing (ILIDS) is a valuable technique to measure the size of droplets or bubbles in a flow [1-4]. It can be extended to the measurement of irregular particles (Interferometric Particle Imaging: IPI). The interferometric image of irregularly shaped particles under laser illumination is a speckle image [5-6]. In the case of ice crystals, sand grains, coal particles, volcanic ash, or glass particles, experimental studies have shown that these interferometric images are directly related to the shape of the particles [6-10]. Thus, the 2D Fourier transform of the interferometric image is linked to the 2D autocorrelation of the particle contour [7]. Validated in many experimental cases, this property can be deduced from a scalar model that assimilates irregular particles to a set of emitting points randomly distributed in the particle contour [6,7]. This property has made it possible to create measuring devices capable of determining the size and certain morphological characteristics of particles whose dimensions can range from a few tens of micrometers to several millimeters [7-11].

However, one may wonder whether this property linking interferometric image and particle shape can be extended to domains where particles are smaller. It seems less likely that a particle of about fifteen micrometers or less, without particular roughness, but simply having an irregular shape can still be assimilated to a large number of emitting points located within its contour. In the case of ice crystals in particular, whose characterization in the atmosphere is an issue for meteorology and aviation safety, the experimental measurement of the shape of crystals of a few micrometers or less remains a significant challenge [13-16]. For these size scales, measurement techniques allowing the visualization of the shape of particles in real flow, and having a measurement field of a few square centimeters, are rare. On the one hand, the relation between spatial frequencies of scattering images and particle size is universal and forms a basis of a multitude of, so-called, spectral methods of single-particle sizing which work down to 1-2 µm [17]. On the other hand, determining the particle shape, or its projection as in interferometric imaging, is inherently more complicated. Hence, its extension to smaller sizes is a priori questionable.

While Lorenz-Mie theory can predict the interferometric image of spherical droplets rigorously, other descriptions are needed to accurately calculate the interferometric image of particles with shapes as complex as ice crystals, under laser illumination. The discrete dipole approximation (DDA) [18,19] is one of the rigorous methods, that can handle particles with sizes up to tens of micrometers, given sufficient computing resources. Another challenge is to be able to describe the shape of ice crystals as accurately as possible. They were obtained using the phase-field model (PFM) developed by Demange et al. [20,21]. From a general perspective, the phase-field model allows to simulate various phase transitions at mesoscale, including solid-solid transformations [22] and solidification/condensation processes [23], which can result into complex morphological changes, such as particle splitting [24], faceting [25], spontaneous or induced side-branching [26], etc. In particular, in [20,21], a new PFM in three dimensions was proposed, that models the ice vapor phase transition, through both anisotropic water molecules attachment and condensation. It has been demonstrated to reproduce the 3D growth dynamics of the most challenging morphologies of snowflakes from the Nakaya diagram [27]. that successfully reproduced manifold faceted morphologies of snowflakes in three dimensions from the Nakaya diagram, including flat fern dendritic ice crystals, needles, faceted sectored planes, and even capped columns.

In the present study, we use DDA to calculate the interferometric images of ice particles of various shapes, modeled by phase fields. Based on this synthetic experiment, we analyze the applicability of IPI to particle sizes down to a few micrometers. Specifically, we compare the 2D Fourier transform of the interferometric images to the 2D autocorrelation of the particle shape to validate the existing measurement principle. We also discuss the compromise between a wide viewing angle, required for smaller sizes, and variations of apparent particle geometry with this angle. Such validated synthetic experiment also provides unique opportunities for various data-intensive deep learning tools [28-31].

1/ Configuration under study

The configuration considered in this study is presented in figure 1. An ice particle is centered in a cartesian coordinate system $(x, y, z)$. The direction of propagation of the incident light defines the z axis, in the forward direction. Incident light is a monochromatic unpolarized plane wave. The scattering direction is given by axis $z'$. $(Oz, Oz')$ define the scattering plane. A CCD sensor is assumed to be located around axis $z'$ (see figure 1). Its center is at distance $B$ of the center of the particle. The axes of the sensor are noted $x'$ and $y'$. $(x', y', z')$ defines an orthonormal coordinate system in the image sensor's frame, while $(x, y, z)$ refers to the particle's frame. Different shapes of ice particles will be considered. They are all calculated rigorously using phase-field modeling.

The light scattering behavior of each modeled ice particle is calculated using DDA, as implemented in the open-source ADDA code [32]. We use ADDA v.1.5.0-alpha3 with filtered-coupled-dipoles formulation of DDA [33] and initialization of the iterative solver through the Wentzel–Kramers–Brillouin approximation [34], which were found helpful for the considered range of sizes and refractive index. Other simulation parameters are set to the default values, with exception to the discretization level that is discussed below. Calculations deliver the Mueller scattering matrix. Simulations will consider unpolarized light and, thus, the $S_{11}$ parameter associated to each $(\theta, \phi)$ pair associated to each pixel of the sensor (see Fig. 1). Specifically, the detector reading (up to a common normalization factor) is given by $S_{11}(\theta, \phi) \cos^3 \beta$, where $\beta$ is the angle between the directions to the pixel and to the detector center. The third power comes from differences in distances and orientation of pixel surface relative to the scattered light. Simulations are performed for particles as small as a few micrometers (between 5 and 25 micrometers).

In simulations, the wavelength is $\lambda = 600$ nm, ice refractive index is 1.33, and the distance $B$ is set to 4 cm. For each shape, we will consider different positions of the image sensor, i.e. different pairs of parameters $(\theta_m, \phi_m)$. These two values refer to the pixel in the middle of the sensor. For a given position of the sensor, each pixel of the sensor will be associated with its corresponding $(\theta, \phi)$ pair.

According to previous studies, the highest spatial frequency present in the speckle pattern should be observed to be $\Delta/(\lambda B)$, where $\Delta$ is the largest dimension of the particle [6]. For particle sizes between 5 and 25 micrometers, it means that the typical size of the speckles in the speckle patterns will be in the range [0.96mm;4.8mm]. In order to observe enough speckles per interferometric image, we will thus simulate sensors dimensions of $3 \times 3$ cm and $4 \times 4$ cm in the case of particles with the largest sizes of approximately 20 and 10 µm, respectively (as specified in later sections). In our simulations, we have considered different forward scattering patterns corresponding to the pairs $(\theta_m = 55°, \phi_m = 0°)$, $(\theta_m = 90°, \phi_m = 0°)$, $(\theta_m = 55°, \phi_m = 90°)$, $(\theta_m = 40°, \phi_m = 35°)$, and for comparison backward scattering patterns corresponding to $(\theta_m = 135°, \phi_m = 0°)$. In addition, the particle can be rotated around axis $y$ (Euler angle $\psi = 0°, 55°$ or $-22.5°$ for different simulations).

Next section is devoted to the presentation of the developed analysis. We discuss progressively different observed effects and the tools developed for proper analysis. Only one shape of particle is considered in this first section: a stellar dendrite ice particle (see Fig. 2). Each presented case is characterized by the pair of angles $(\theta_m, \phi_m)$ that give the position of the center of the imaging sensor's plane, and the angle $\psi$ that gives the orientation of the particle. In further sections, we present the results obtained with other particle shapes and sizes to generalize the study.

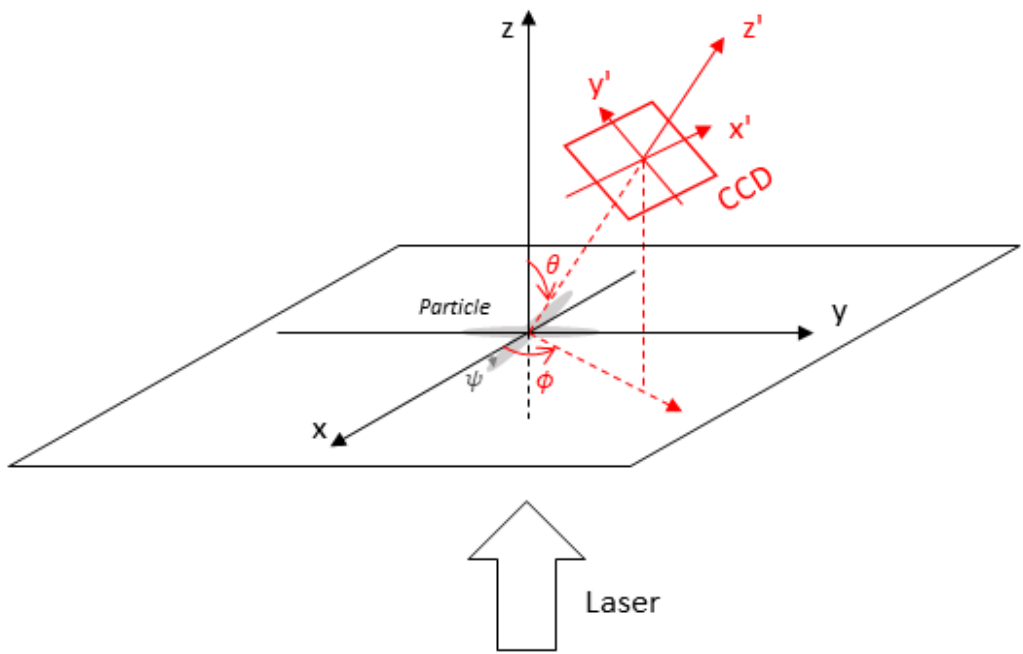

Figure 1 : Configuration under investigation.

## 2/ Detailed analysis for a stellar dendrite ice crystal

### 2.1/ Windowing the interferometric image

Figure 2 shows the first ice particle to be considered: a stellar dendrite. Figure 3(a) shows its projection in its own reference frame ($x$ and $y$ axes of Fig. 1). The particle's orientation is given by $\psi = 0°$. The sensor's position corresponds to a pair $(\theta_m = 55°, \phi_m = 0°)$. Figure 3(b) shows the particle as it appears from the center of the image's sensor, i.e. in the frame of the $x'$ and $y'$ axes of Fig. 1. The length of the particle along the $x$-axis is 26.7 µm in this case, while the volume-equivalent size parameter is 46.9 . Note that this crystal is flattened, with a small width along the z-axis (aspect ratio of 15.1). The 2D-autocorrelation of this sensor-based projection has been calculated and binarized in figure 3(c). The contour of this 2D-autocorrelation is reported in red. The aim of the following light scattering simulations is to know whether the 2D-Fourier transform of the interferometric image of the particle will correspond to this Fig. 3(c).

Figure 4(a) shows the interferometric image calculated with DDA, using 15 dipoles per wavelength, which is slightly larger than the rule-of-thumb for such refractive index [18]. Thus, we expect fine accuracy, i.e. DDA can be considered a rigorous method, especially in comparison with approximations considered below. Note, however, that directly testing the accuracy by refining discretization is not trivial, since the used discretization exactly corresponds to the one used in the phase-field modelling. The image is approximately symmetric, according to the quasi-symmetry of the particle for the considered angle of view (the shape obtained from phase field modelling is actually not perfectly symmetric).

Figure 4(b) shows the modulus of the 2D-Fourier Transform (2DFT) of the interferometric pattern. The scaling factor $\lambda B$ has been applied to the axes of the spatial frequencies $u$ and $v$, in order to enable quantitative comparisons. On these figures we also show in red the contour of the 2D-autocorrelation of the particle's shape, already plotted in Fig. 3(c). Results in Fig. 4(b) show that the spatial frequency spectrum is encircled by the red contour. Unfortunately, a significant cross-like structure is observed, in particular intense vertical lines in Fig. 4(b), due to the aliasing inherent to the numerical 2DFT operation. More specifically, the interferometric pattern shows much higher intensities on the top of the sensor than that on the bottom. In such conditions, the application of a window (to impose periodic limit conditions) before 2DFT operation eliminates the aliasing. We have chosen the simplest window, a multiplication of the two cosine functions that vanish on the borders of the image (for both $x$ and $y$ axes). Fig. 4(c) shows the windowed interferometric pattern, while fig. 4(d) shows the modulus of its 2DFT. The cross artefact is no more noticeable, resulting in a relatively good correspondence between the spatial frequencies spectrum of the interferometric pattern and the 2D-autocorrelation of the particle's shape. Note that during the 2DFT operation, we always observe a very intense peak in the center of the spectrum. It is attenuated on Fig. 4(b), (d) in order to better evidence the surrounding spatial frequencies spectrum (see the central yellow square).

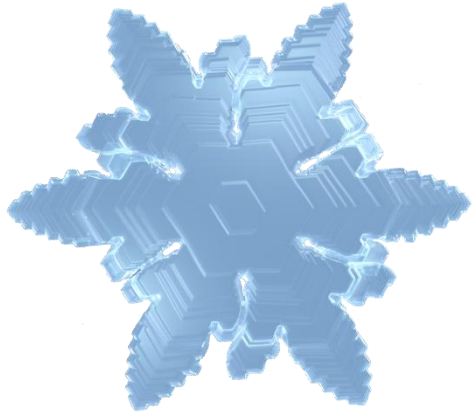

Figure 2 : Main face of a stellar dendrite.

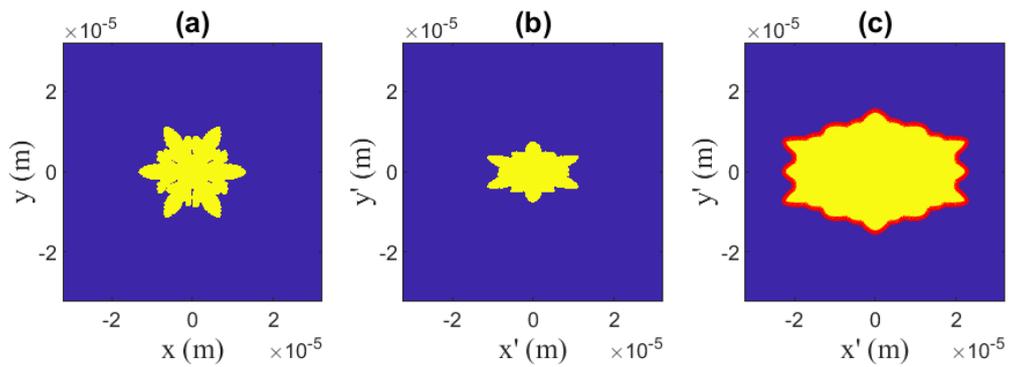

Figure 3 : projection of a stellar dendrite in the particle's frame (a) and as viewed from the sensor (b). Part (c) depicts the binarized 2D-autocorrelation of (b).

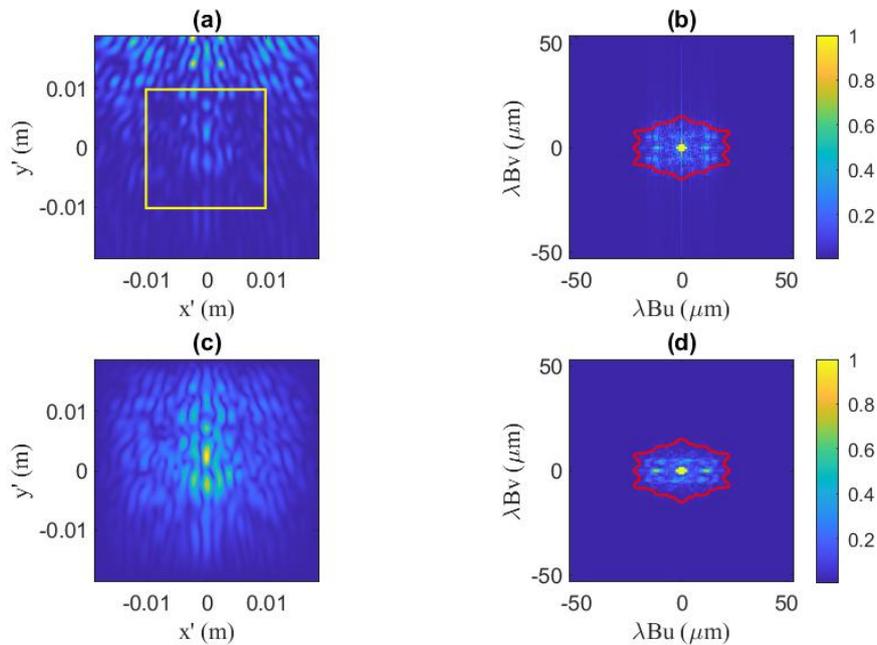

Figure 4 : interferometric image computed with DDA (a) and its 2D-Fourier transform in (b); windowed interferometric image (c) and its 2D-Fourier transform (d). The red contour in (b) and (d) corresponds to the 2D-autocorrelation of the particle's projection. The yellow box in (a) shows the selected part of interferometric images that will be analyzed in next section 3.

2.2/ Huygens-Fresnel conditions

Assuming that an irregular particle can be represented by an ensemble of coherent point emitters randomly located in the contour of the particle's shape, and with random phase, the relation that links the 2DFT of the interferometric pattern and the 2D-autocorrelation of the particle's face is obtained using Fresnel's kernel, i.e. in Fresnel conditions. As we consider a relatively large sensor in order to receive light from large ranges of $\theta$ and $\phi$ values, Fresnel conditions could be not fully satisfied. To test this effect, we have modeled the particle by an ensemble of 609 point emitters. Its interferometric image has then been calculated using either Huygens-Fresnel or Luneburg's kernel (the latter respects the first Rayleigh-Sommerfeld integral) [35]. We have then applied the same window as for DDA patterns, the results are presented in Figures 5(a) and 5(c), respectively. The moduli of their 2D-Fourier transforms are presented in (b) and (d), respectively. We can observe that the spatial frequencies spectrum is well encircled by the contour of the 2D-autocorrelation of the particle's shape (in red), as expected in Fresnel conditions. The difference between the two windowed patterns (both normalized to the maximum value of the intensity) is presented in figure 6. We can observe that after the window operation, the patterns obtained using both models are very similar, with low differences, indicating that they mostly satisfy Huygens-Fresnel conditions. Thus, the window operation happens to present a second advantage: in addition to eliminating the aliasing artefacts, it suppresses the border part of the image – the one that is most questionable with respect to the Fresnel conditions.

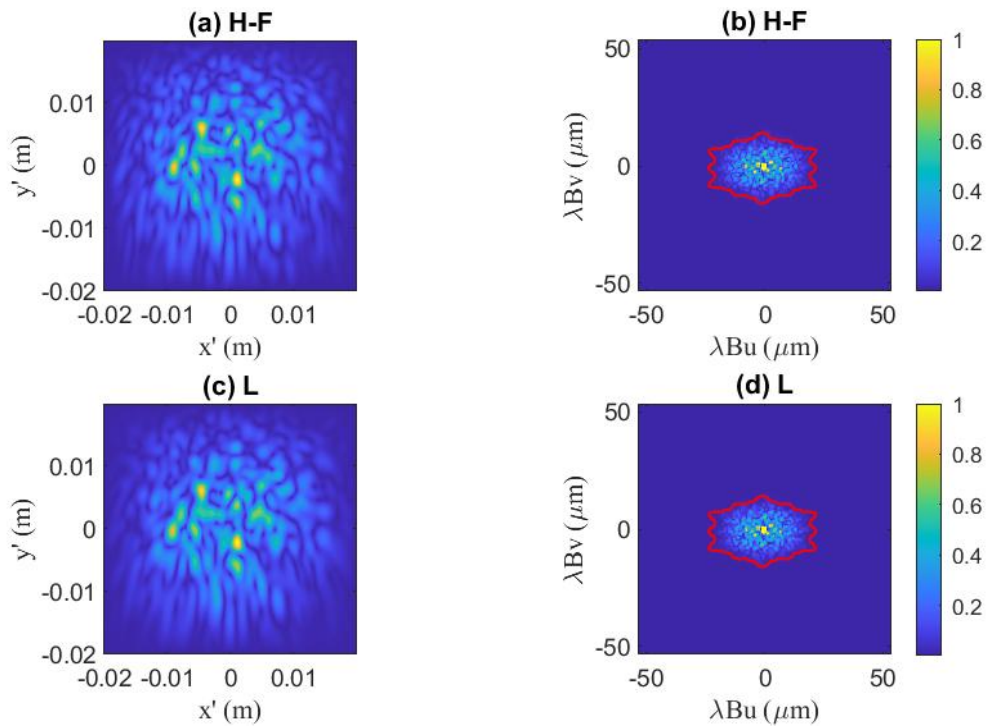

Figure 5 : windowed interferometric images of an ensemble of point emitters randomly located in the contour of the stellar dendrite particle using Huygens-Fresnel (a) or Luneburg's integral (c), and their respective 2D-Fourier transforms (b) and (d). The red contour corresponds to the 2D-autocorrelation of the particle's shape.

Brunel, Demange, Patte, Yurkin

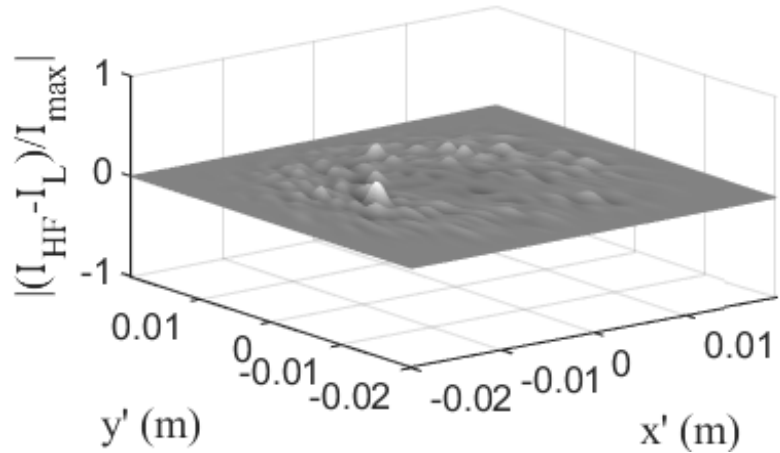

Figure 6 : Normalized difference of windowed interferometric images obtained using Huygens-Fresnel and Luneburg's integrals (Fig. 5(a) and (b), respectively).

It is important to remember that the two interferometric patterns calculated using Huygens-Fresnel and Luneburg integrals do not present the real patterns, in contrast to the DDA ones. In particular, these two patterns do not present any symmetry because the point emitters in the contour of the particle are chosen randomly and not located symmetrically. However, by design, the 2DFT of these patterns contains the signature of the particle shapes, i.e. its 2D-autocorrelation. The same property is not immediately obvious for DDA, but is confirmed throughout this paper.

2.3/ Diversity of the angles of view on a same image

Figure 7 shows the interferometric image obtained using ADDA calculation in another case. The particle is the same as in previous section but its orientation is now given by $\psi = -22.5°$. The position of the center of the sensor is given by the pair $(\theta_m = 55°, \phi_m = 90°)$. Four different windows of the pattern have been selected on figure 7. The borders of the main image is removed from consideration to stay within the validity of Fresnel conditions. The four framed speckle patterns show quantitatively different morphological properties: elongated speckles in the red window, and more circular speckles in the yellow one. The framed patterns have been further windowed individually, as explained in previous section, and then 2D-Fourier-transformed. Figures 8(a-d) show the obtained transforms together with the contours of the 2D-autocorrelations of the particle projection as observed from the center of each different sub-image (the line color corresponds to that of the original frames). There is significant difference between the contours due to the variation of the view angle, but this variation is well followed by the retrieved spatial spectrum. For all sub-images, there is a good agreement between the two.

Brunel, Demange, Patte, Yurkin

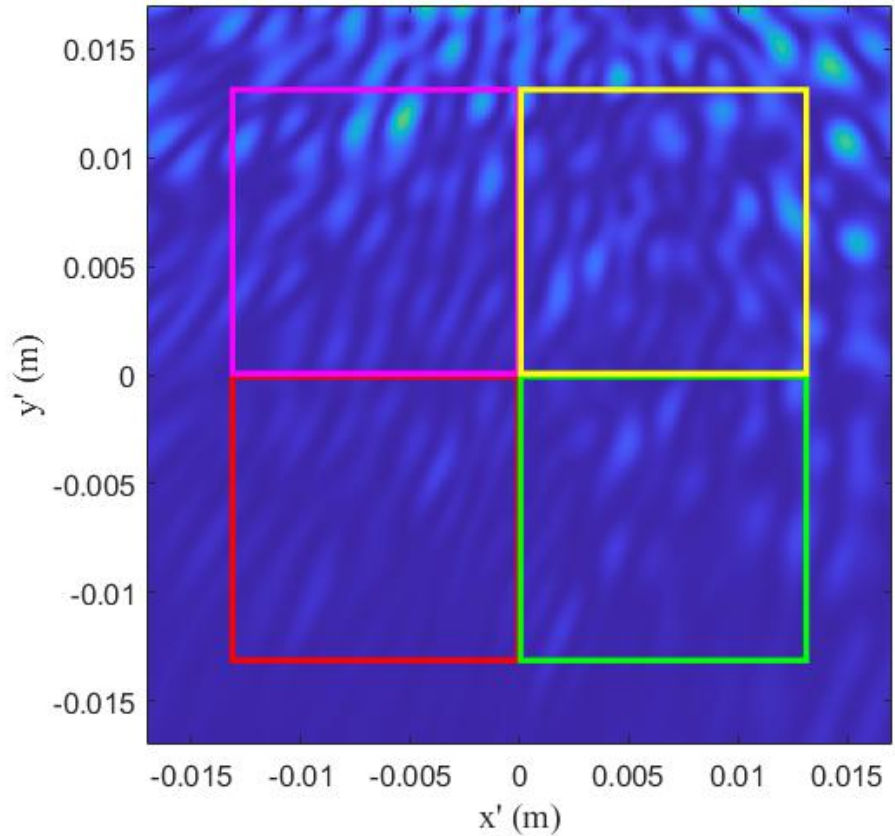

Figure 7 : interferometric image of an ice particle whose orientation is given by $\psi = -22.5°$. The position of the sensor center is given by the pair ($\theta_m = 55°, \phi_m = 90°$).

This case illustrates the fact that it is actually not obvious to define a contour of the particle's shape considering a sensor with relatively large angular extent (which is necessary in the case of small particles). Each pixel of the sensor observes the particle from its own angle of view. This effect becomes significant when the scattering angle $\theta$ varies by 40 or more degrees between the top and the bottom of the original image of figure 7 (idem for $\phi$). It is then interesting to analyze different parts of the whole image separately as done in figure 8. However, this can only be done if the selected sub-images contain at least a few speckles, i.e. if the particle is not too small.

Note that in the case of quasi-planar ice particles, the significance of this effect strongly depends on the particle orientation. In particular, the difference between Fourier transforms of sub-images is the largest when the edge of the particle is observed, and the least – for the face. Moreover, this effect was already visible in figure 4(a): speckles turn into longer filaments at the bottom of the image (i.e., for view angles corresponding to the particle edge).

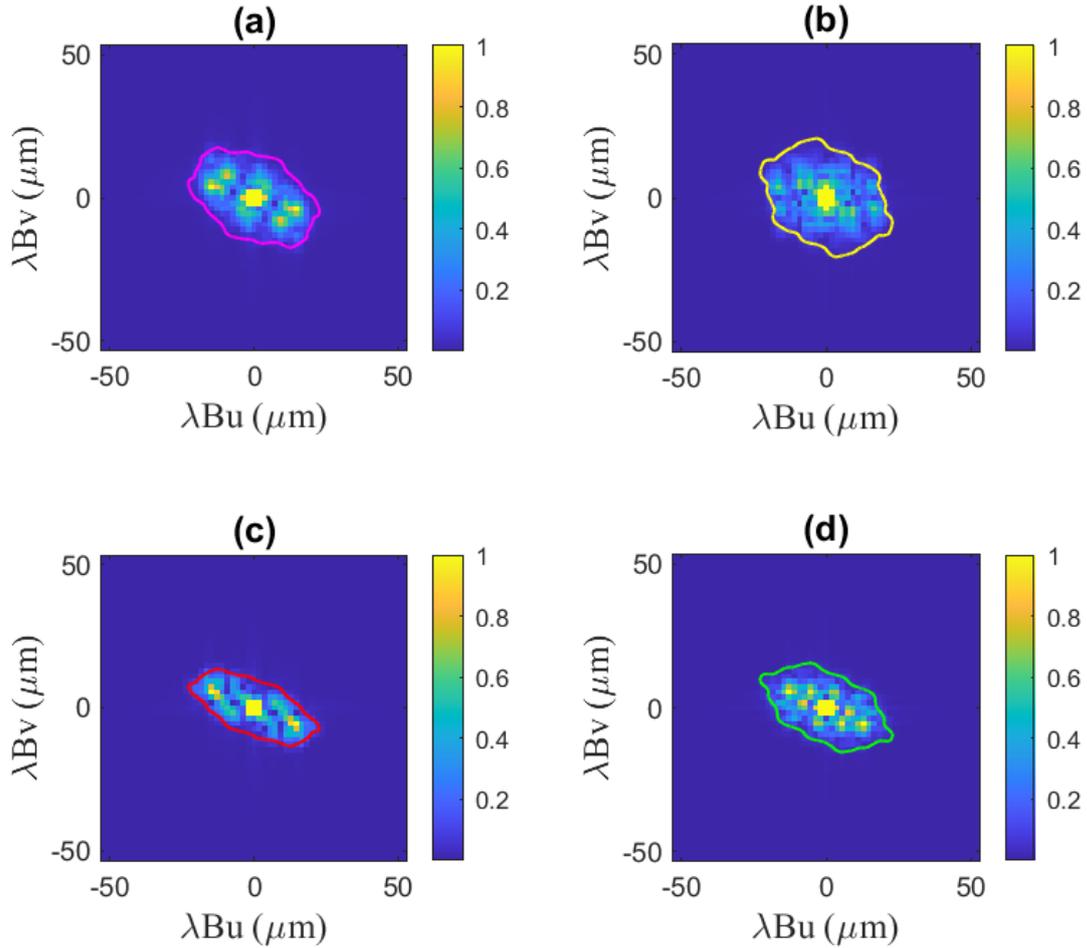

Figure 8 : 2D-Fourier transforms of the four partial images of figure 7 together with contours of the 2D-autocorrelation of the particle's shape for each angle of view. Colors of the contours correspond to colors of image frames.

In addition to this difference of viewing angles, the further from the center of the sensor the pixels are, the less their plane is perpendicular to the rays scattered by the particle. This effect induces distortion of the interferometric image, as illustrated in Figure 9 where the particle is centered on $O$ while the sensor is centered on $O'$. To study the influence of this distortion, we have calculated separately the interferometric images of the four selected portions of the sensor of figure 7 (magenta, yellow, red, green boxes). In these new calculations (using ADDA code), each portion of the sensor has been rotated to be perpendicular to the ray scattered in the direction of the center of the portion. This is illustrated on figure 9, reduced to a 2D-case, with two detector centers $O'_1$ and $O'_2$ that are rotated by angles $\delta\varphi_1$ and $\delta\varphi_2$ respectively.

The four interferometric sub-images are reassembled and shown in figure A1 of appendix A. These four sub-images have then been analyzed separately as if they were delivered by four independent sensors: the four sub-patterns are windowed and 2D-Fourier transformed separately, and compared to the contours of the 2D-autocorrelations of particle's projection observed from the center of the corresponding sub-image. The scaling parameter is now $\lambda B_1$ where distance $B_1$ is calculated from the position of each subimage's center (see figure 9). Results are presented in figure 10 – the correspondence between 2D-autocorrelations of particle's contours and 2D-Fourier transforms of the patterns is comparable to that in figure 8.

In summary, the angle of view of the particle from the pixels under consideration appears to be a very important parameter in the pattern's analysis (figures 7,8). By contrast, the additional distortion of the image (figure 10) appears to have only minor influence; still, it should be considered to enhance the interpretation. Note also, that such virtual rotation of the detector can be performed for any measured signal based purely on geometric considerations and interpolation.

In order the avoid the difficulties described in this section, in the following sections we focus on the center part of the original pattern (yellow box of figure 4 for example).

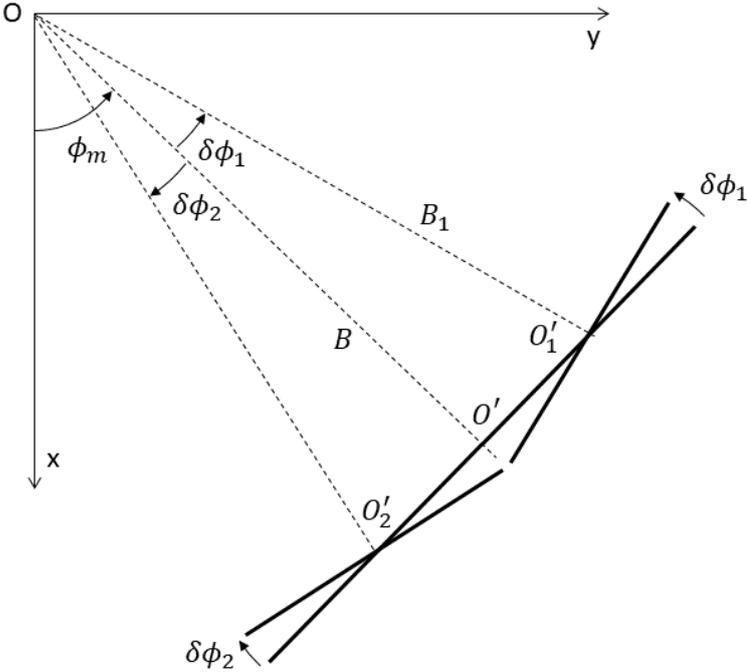

Figure 9 : Reorientation of a portion of the large sensor for the correction of distorted images.

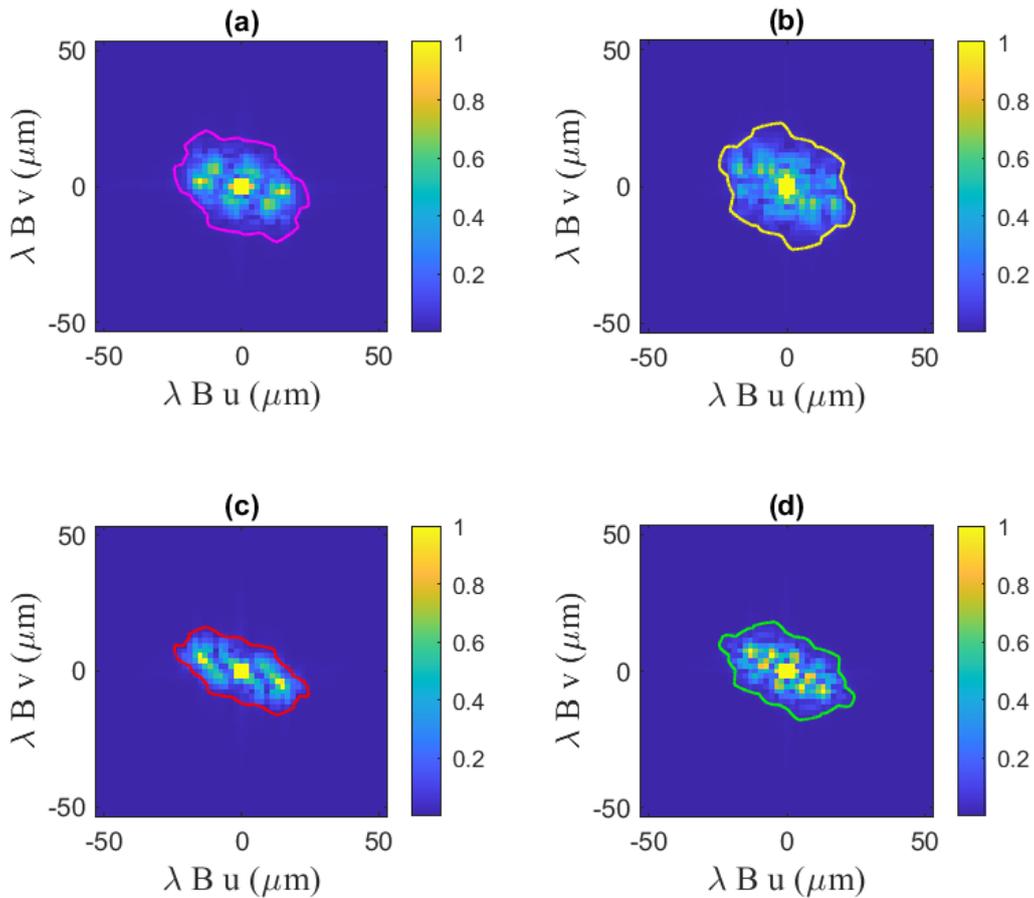

Figure 10 : 2D-Fourier transforms of the four partial images of figure 10 together with contours of the 2D-autocorrelation of the particle's shape for each angle of view.

2.4/ Procedure for automated analysis

Let us summarize the procedure that will be used in the following section.

- Using ADDA code, the interferometric image of an ice particle of given shape and orientation is calculated for a given position of the sensor (corresponding to a pair $(\theta_m, \phi_m)$).
- We select the center part of this image to stay in Fresnel conditions (yellow box of figure 4a). It should be limited in its angular range, but large enough to contain at least a few speckles. This allows us to use only a single image in contrast to more complicated procedure discussed in the previous section (figures 7-10).
- This central part of the interferometric image is windowed to eliminate the cross-structure due the strong discontinuities across the frame border.
- This windowed selection is 2D-Fourier transformed (applying scaling factor $\lambda B$) and compared to the contour of the 2D-autocorrelation of the particle's shape.

Keeping in mind the effects discussed in previous sections, we have developed a numerical procedure to estimate the contour of the 2D-autocorrelation of the particle's shape: each interferometric image or sub-image that will be analyzed is virtually divided into 49 equal parts (from the top left of the selected image to the bottom right). We calculate then the shape of the particle viewed from the center of each of these 49 partial images. The final 2D-autocorrelation of the particle's shape (whose contour will be reported) is the sum of these 49 2D-autocorrelations. This choice of 49 angles of view is a compromise to reduce computation time (we could have indeed chosen to consider all angles of view associated to all pixels of the image). Note that this division in 49 partial images is only used to estimate the contour of the 2D-autocorrelation of the particle's shape.

Potentially, more information about the particle shape can be retrieved by tomographic like reconstruction (rigorously accounting for angular position of each detector pixel). However, we deliberately limit ourselves to the simple 2D procedure.

3/ Automated analysis for other particle's shapes and different angles of view

We have performed the analysis for other particle's shapes, sizes and orientations, and different angles of view, using the procedure postulated in the end of the previous section. Figure 11 shows the different ice crystal shapes that we considered. From left to right, the aspect ratios of these crystals are 5.9 for the simple star, 7.8 for the stellar crystal, and 10 for the sectored plate. It was 15.1 for the stellar dendrite of figure 2.

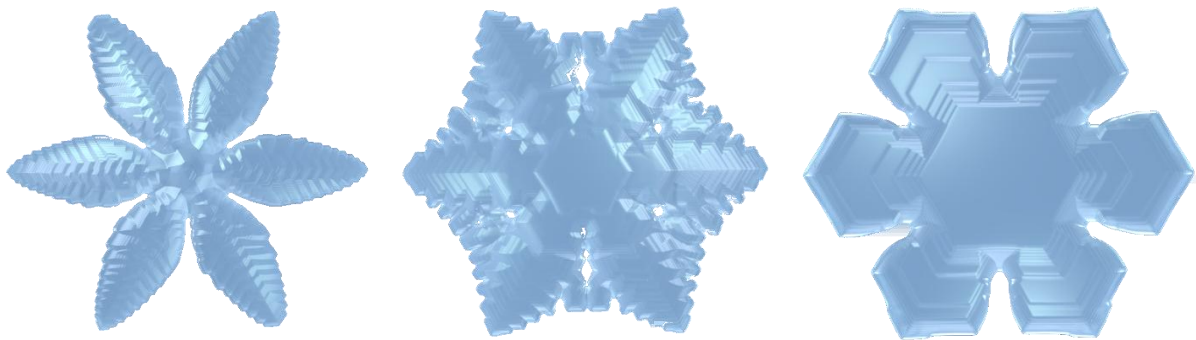

Figure 11 : Main faces of other crystal shapes considered in this study, denoted as simple star, stellar crystal, sectored plate from left to right.

For example, Figure 12(a) shows a stellar crystal, as it appears from the center of the sensor positioned at $(\theta_m = 40°, \phi_m = 35°)$, while the crystal is parallel to the $(x,y)$-plane (i.e., all Euler angles equal to 0). Figure 12(b) shows the 2DFT of the windowed interferometric pattern in comparison with the red contour of the 2D-autocorrelation of the particle's shape. There is a qualitative agreement between the two, but the sensor-derived spatial spectrum does not have a well-defined boundary. Figure 12(b) shows that the spatial spectrum presents domains with high intensity, and also domains with low intensity where rapid variations of the intensity level are observed (one example is indicated by the arrow). Thus, we developed a binarization process to define such a boundary. This process combines a threshold binarization of the spatial spectrum, and a threshold binarization of the gradient of the spatial spectrum to also consider small-intensity domains with strong variations.

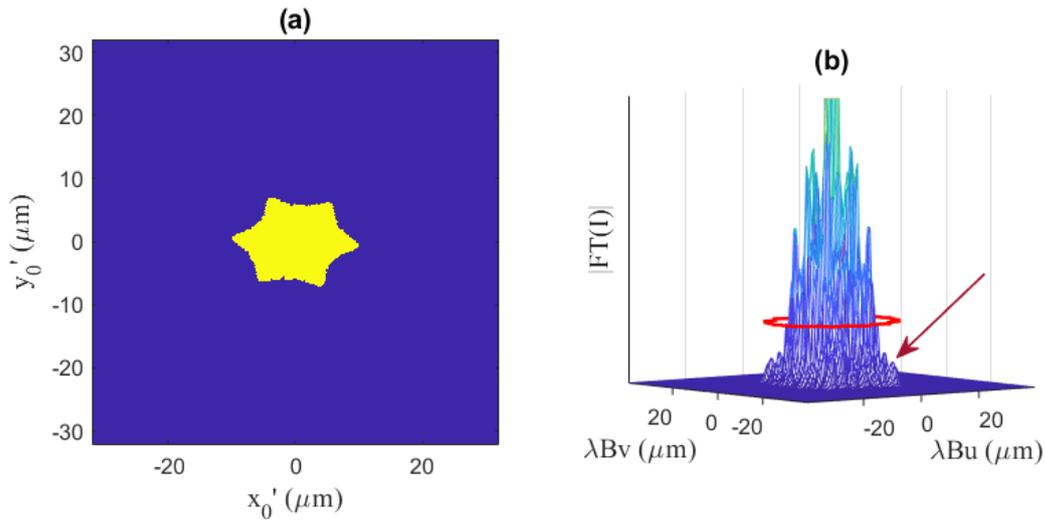

Figure 12 : Stellar particle for sensor position ($\theta_m = 40°, \phi_m = 35°$) (a) and the 2DFT of the windowed interferometric pattern (b) (the red contour is the 2D-autocorrelation of the particle's shape, placed at arbitrary vertical position for better visibility).

The following figures present the results obtained in different forward scattering configurations, corresponding to the pairs ($\theta_m = 55°, \phi_m = 0°$), ($\theta_m = 55°, \phi_m = 90°$), and ($\theta_m = 40°, \phi_m = 35°$). In addition, two orientations of each particle will be considered, given by $\psi = 55°$ or $\psi = -22.5°$. We, thus, test 6 cases for each particle (3 columns by 2 rows in the presented figures). The size of the central part of the interferometric pattern that is selected before 2DFT is: $2.3 \times 2.3$ cm. As the distance between the particle and the sensor is $B = 4$ cm, the variations of angles $\theta$ and $\phi$ from their mean values $\theta_m$ and $\phi_m$ is smaller than $\pm 16°$. The selected patterns are then analyzed as described at the end of section 2.4. Figures 13–16 show the results obtained for the particles of figures 12 and 2, respectively: simple star, stellar crystal, sectored plate, and stellar dendrite. Each sub-plot compares the contours of the spatial frequencies' spectra (scaled by $\lambda B$) to that of the 2D-autocorrelations of the particles' shapes. The binarization procedures, including the thresholds, are the same in all presented cases to test the feasibility of an automated analysis.

We present only the results obtained with the smallest particles, whose largest dimensions are 10 µm, 9.9 µm, 6.9 µm, and 13.3 µm for figures 13 to 16, respectively (the results obtained with twice larger particles can be found in appendix B). The corresponding volume-equivalent size parameters are 21.4, 21, 14.9, and 23.5. For convenience (at the cost of wasting some computational resources), we used the same input shape files and adjusted the size by halving the dipole size, i.e., increasing the number of dipoles per wavelength to 30. In most cases, there is a good qualitative agreement between the contours. But the results highlight the difficulty of universal automatic binarization. Figures 13(c) and 16 show cases where the threshold used for 2DFT binarization seems to be too high. By contrast, on figures 13(d), 15(d), the contours of the 2DFT remain slightly wider. These last cases correspond to the observation of a small highly-oblate particle from its edge.

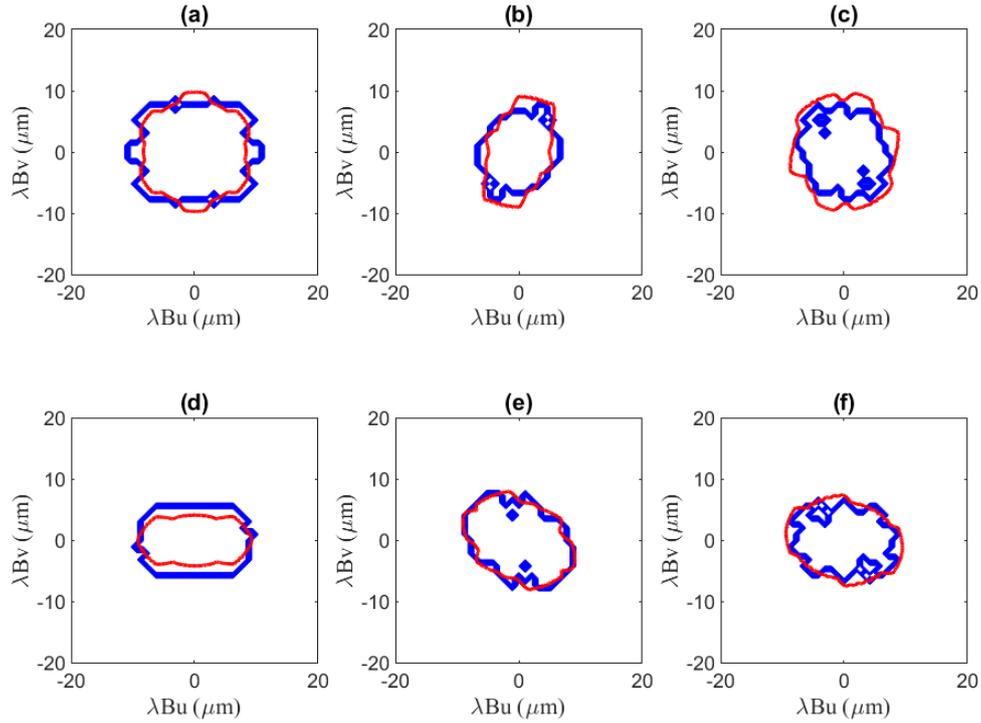

Figure 13: Contour of the 2D-autocorrelation of the particle's shape (in red) and contour of the 2DFT of the interferometric pattern (in blue) for a simple star ice particle. $\theta_m = 55°, \phi_m = 0°$ in (a) and (d); $\theta_m = 55°, \phi_m = 90°$ in (b) and (e); $\theta_m = 40°, \phi_m = 35°$ in (c) and (f). The particle's orientation is given by $\psi = 55°$ in (a-c) and $\psi = -22.5°$ (d-f).

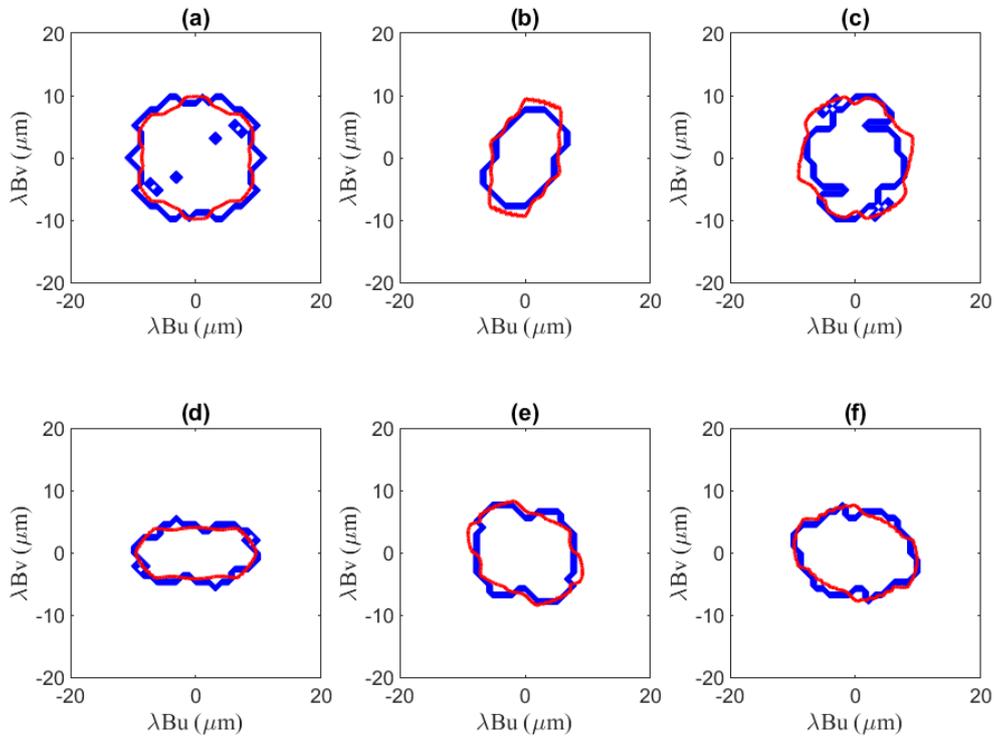

Figure 14: Contour of the 2D-autocorrelation of the particle's shape (in red) and contour of the 2DFT of the interferometric pattern (in blue) for a stellar ice particle. $\theta_m = 55°, \phi_m = 0°$ in (a) and (d); $\theta_m = 55°, \phi_m = 90°$ in (b) and (e); $\theta_m = 40°, \phi_m = 35°$ in (c) and (f). The particle's orientation is given by $\psi = 55°$ in (a-c) and $\psi = -22.5°$ (d-f).

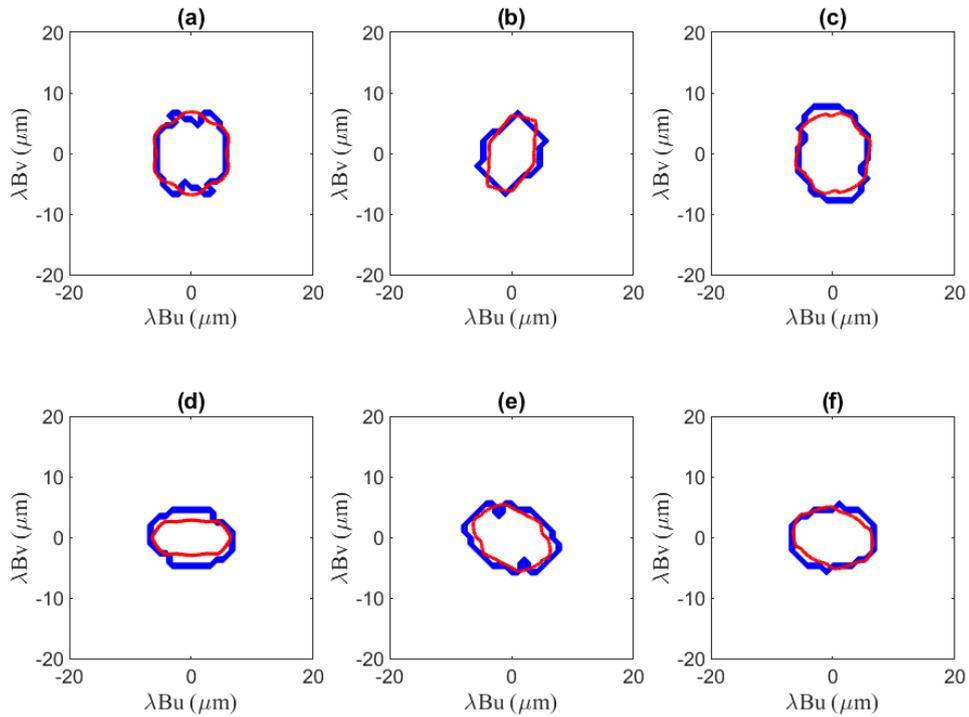

Figure 15: Contour of the 2D-autocorrelation of the particle's shape (in red) and contour of the 2DFT of the interferometric pattern (in blue) for a sectored plate. $\theta_m = 55°, \phi_m = 0°$ in (a) and (d); $\theta_m = 55°, \phi_m = 90°$ in (b) and (e); $\theta_m = 40°, \phi_m = 35°$ in (c) and (f). The particle's orientation is given by $\psi = 55°$ in (a-c) and $\psi = -22.5°$ (d-f).

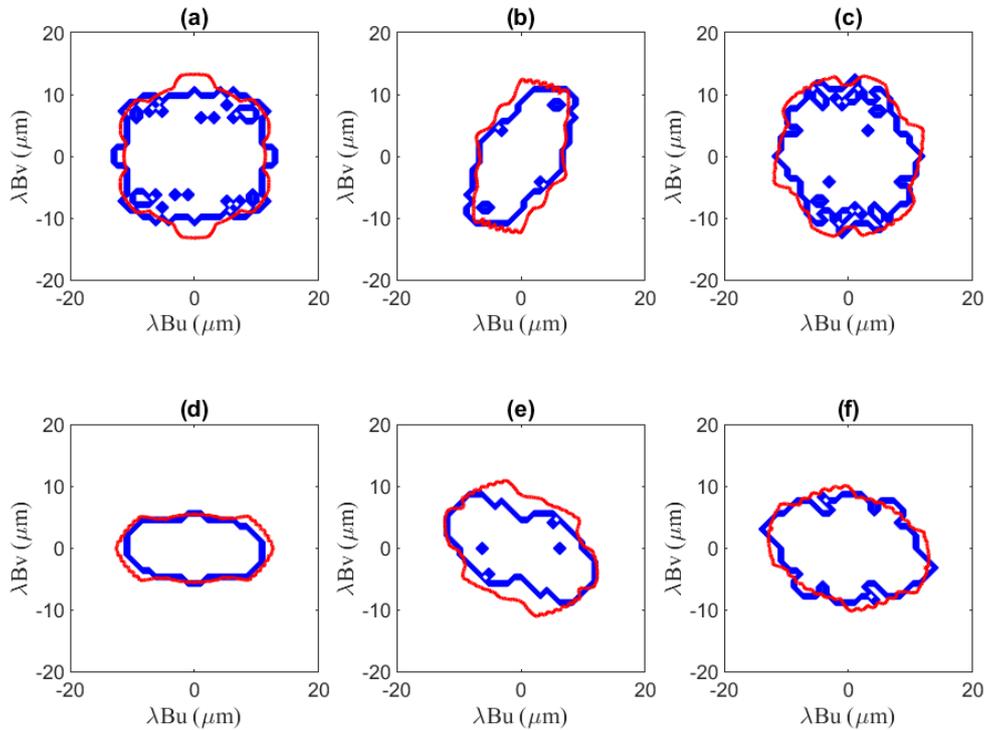

Figure 16: Contour of the 2D-autocorrelation of the particle's shape (in red) and contour of the 2DFT of the interferometric pattern (in blue) for a stellar dendrite. $\theta_m = 55°, \phi_m = 0°$ in (a) and (d); $\theta_m = 55°, \phi_m = 90°$ in (b) and (e); $\theta_m = 40°, \phi_m = 35°$ in (c) and (f). The particle's orientation is given by $\psi = 55°$ in (a-c) and $\psi = -22.5°$ (d-f).

Brunel, Demange, Patte, Yurkin

Let us further discuss the low-size limit of IPI. For instance, two complete fringes are needed to measure the size of a spherical droplet in ILIDS [1]. But the interferometric pattern is then a regular two-wave interference pattern, rather than a random speckle as for ice crystals. Assuming that the speckle pattern observed on a sensor (of dimensions $D \times D$) is composed of $2^2$ regularly spaced bright speckles, the corresponding spatial frequency measured on the sensor is given by $f = 2/D$ . As mentioned in the Introduction, the highest frequency expected in the IPI speckle pattern is given as $f_{max} = \Delta/(\lambda B)$ where $\Delta$ is the particle's size and $B$ the distance between the particle and the sensor. Equating $f$ and $f_{max}$ we obtain a smallest measurable size $\Delta_{min} = 2.1$ µm for a sensor of dimension $D = 2.3$ cm at distance $B = 4$ cm from the particle (as for figures 13-16).

Unfortunately, this simple relation is not applicable to the case of particles whose aspect ratio is very different from 1. Figure 17 shows the case of a parallelepiped particle with a large aspect ratio and a sensor placed as for the parts (d) of figures 13-16. For point $A_1$ of the sensor, the apparent size of the particle is the transverse thickness $dt$. But for point $A_2$, it is approximately $dl \sin(\alpha + \chi) + dt \cos(\alpha + \chi)$. The maximum angle $\alpha_{max} = \text{atan}(D/(2B))$ associated to point $A_3$ corresponds to the highest observed spatial frequency. This has two important implications. First, for a given particle width, say $dt = 0.5$ µm, the above value of $\Delta_{min}$ leads to the minimum observable lateral dimension $dl = 3.4$ µm. Second, for a given lateral dimension, say $dl = 10$ µm, the size observed from $A_3$ will be 5.2 µm. However, once can consider a smaller sub-image (reducing $\alpha_{max}$) to reduce it in compromise with increasing $\Delta_{min}$. This approach would be similar to processing several sub-images in Section 2.3 and we leave it for a future research.

With the simulated set-up, the apparent sizes of the crystals under study (obtained from the synthetic experiment) are 6.2 µm for the simple star (Fig. 13(d)), 5.2 µm for the stellar crystal (Fig. 14(d)), 5.2 µm for the sectored plate (Fig. 15(d)) and 6.2 µm for the stellar dendrite (Fig. 16(d)). These measurements are relatively close to the apparent shapes sizes deduced from the 2D-autocorrelations (red curves). One additional factor for discrepancy is the used window function. Since the latter is a cosine one, the frequencies present in the patterns are increased by that of the window, corresponding to overestimation of apparent sizes by 0.5 µm.

The above discussion scratches the surface of several issues for application of IPI to high-aspect-ratio particles with the smallest size smaller than or comparable to the wavelength. Further studies should consider much larger number of particle orientations, while interpretation of a whole measurement dataset for an ensemble of particles should account for the fact that similar (but not identical) particles are observed in different orientations.

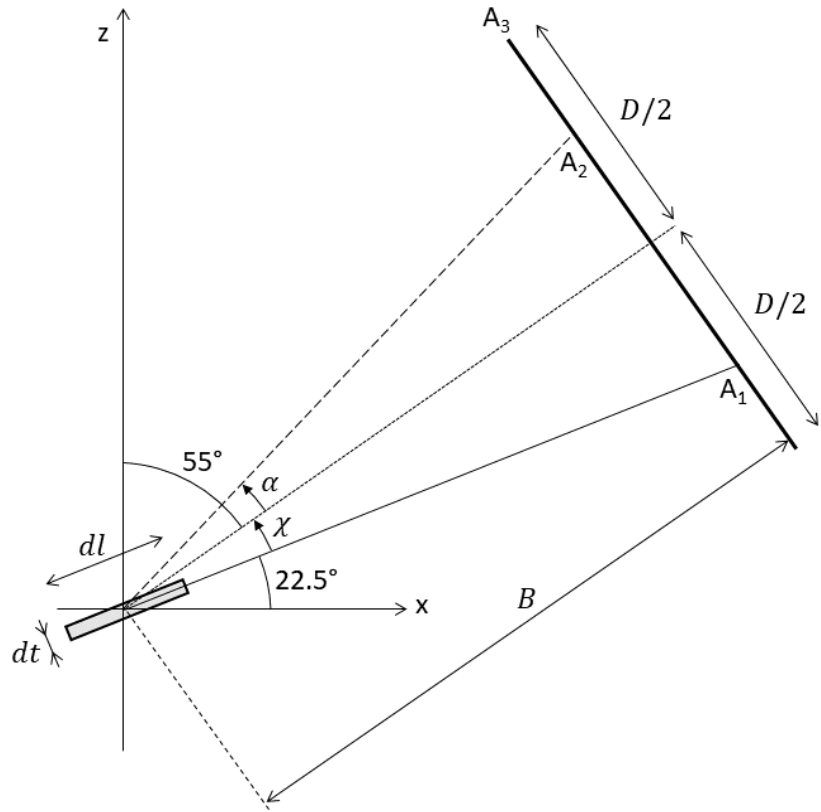

Figure 17: Geometry of the scattering problem for a high-aspect-ratio particle with strong variation of apparent size with viewing angle.

Another particular situation in IPI is the case of neighboring particles whose interferometric images overlap [36]. If the separation between particles is larger than their size, the 2D-autocorreltion of this multi-component object is composed of a central peak and two symmetric peaks that correspond to the cross-correlations between both particles. If the distance is smaller, the 2D-autocorrelation shows a single complex spread spot. If the number of particles increases, the number of cross-correlations between particles increases (6 cross-correlations for a set of 3 well-separated particles, etc…). Figure 18 shows the test case of three sectored plates at normal view angle. Figure 19 presents then the results obtained in the different forward scattering configurations (($\theta_m = 55°, \phi_m = 0°$), ($\theta_m = 55°, \phi_m = 90°$), and ($\theta_m = 40°, \phi_m = 35°$)) for the two orientations of the set of particles ($\psi = 55°$ and $\psi = -22.5°$). These different figures show a very good correspondence between the contours of 2D-autocorrelations and 2DFT. In particular Fig. 19(a) corresponds to a case where all two-particle cross-correlations are well-separated, while fig. 19(d) – where they largely intersect. Figures 19(a,c) deserve extra attention with respect to the small-size limit of the technique. In this case, the size of each of the three similar particles is around 2 µm, and they are oriented so that the different pixels of the sensor observe almost the same shapes. The central red spot is the superposition of the 2D-autocorrelations of the apparent shapes of the three particles. It agrees with 2D-Fourier transform of the pattern, highlighting that the sizing limit of 2 µm is indeed reachable. The other (shifted) peaks, that correspond to the cross-correlations between particles, are clearly separated and have the same width. Their use may become beneficial with further decrease of the particle size. This is the heterodyne principle – spatial frequencies, which are too small to be measured by themselves, can still be measured in combination with a larger shift frequency. We leave, however, the investigation of this capability to future research.

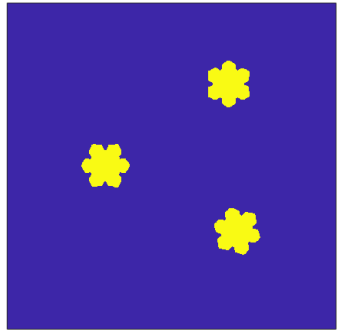

Figure 18: Multi-components object: three separated ice particles.

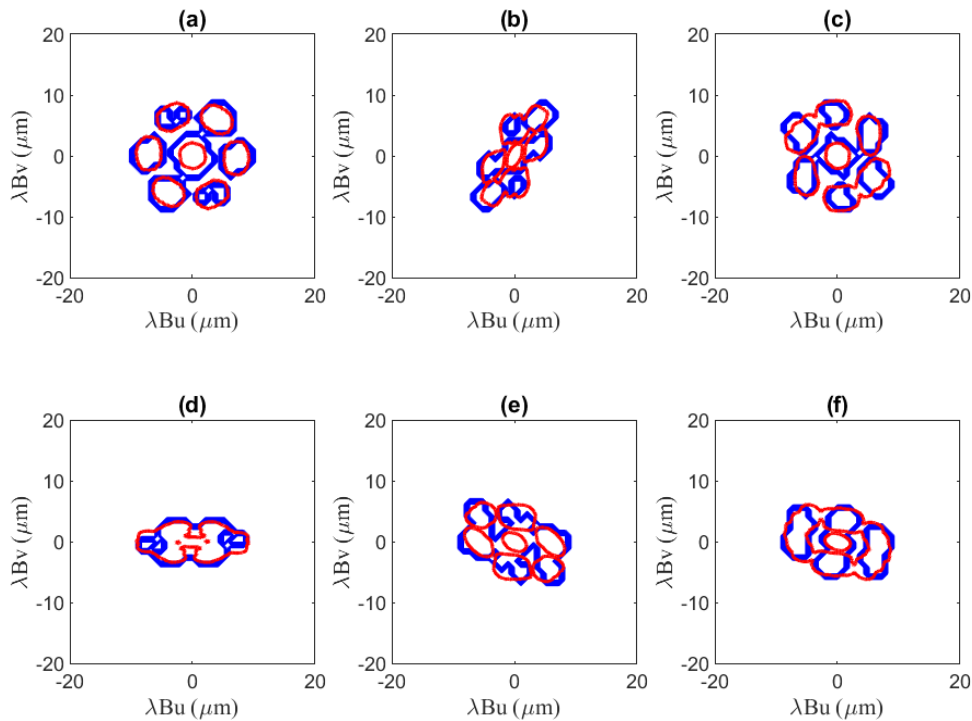

Figure 19: Contour of the 2D-autocorrelation of the multi-components shape (in red) and contour of the 2DFT of the interferometric pattern (in blue) for a set of 3 ice particles. $\theta_m = 55°, \phi_m = 0°$ in (a) and (d); $\theta_m = 55°, \phi_m = 90°$ in (b) and (e); $\theta_m = 40°, \phi_m = 35°$ in (c) and (f). The particle's orientation is given by $\psi = 55°$ in (a-c) and $\psi = -22.5°$ (d-f).

We have performed similar DDA simulations for twice larger particles of the same shapes (simple star, stellar, sectored plate, and stellar dendrite). They are based on the original shape files and close-to-optimal value of the number of dipoles per wavelength (15, the same as in section 2). Very similar results have been obtained. They are reported in the appendix B and not in the main text, as our main focus is on the particles probing the small-size limit of the technique.

We also considered a backward scattering configuration, corresponding to the pair of angles: $\theta_m = 135°, \phi_m = 0°$, for three different orientations of the particles given by $\psi = 55°, -22.5°$. Similar results were obtained when $\psi = 55°$. Results are presented in appendix B for the two ranges of particle sizes: sizes around 10 µm in figure C-1 and sizes twice larger in figure C-2. But the case $\psi =$

Brunel, Demange, Patte, Yurkin

$-22.5°$ led to a different behavior, which is discussed in the following. This case, a combination of $\theta_m = 135°$ and $\psi = -22.5°$, actually corresponds to the observation of the specular reflection. Figures 20 (a-d) show the results obtained for the simple star, stellar, sectored plate and stellar dendrite ice crystal shapes of smaller sizes, i.e. the same one as used in figures (13-16). The processing algorithm, including the binarization thresholds, is the same as well. The contours of the 2DFT of the interferometric patterns (in blue) are in general smaller than the contours of the 2D-autocorrelations (except for the smallest sectored plate particle (Fig. 20(c)).

To explain this behavior, let us look at the full 2DFT profiles, shown in figure 21 for the profiles obtained for the simple star, stellar, sectored plate and stellar dendrite crystals respectively. The red curve report the contours of the 2D-autocorrelation of the corresponding particle's shape for comparison. When observing the specular reflection, the central spatial frequencies form an intense peak that dwarfs the surrounding higher spatial frequencies (in contrast to what was observed on figure 9(b), for example). This effect increases with the particle size (Fig. 21(d)) and makes the result very sensitive to the binarization threshold. While it may be possible to adjust the binarization procedure to perform better on such images, the high dynamic range associated with the specular reflection is even more problematic in experiments. Thus, such orientations remain inaccessible to interferometric imaging.

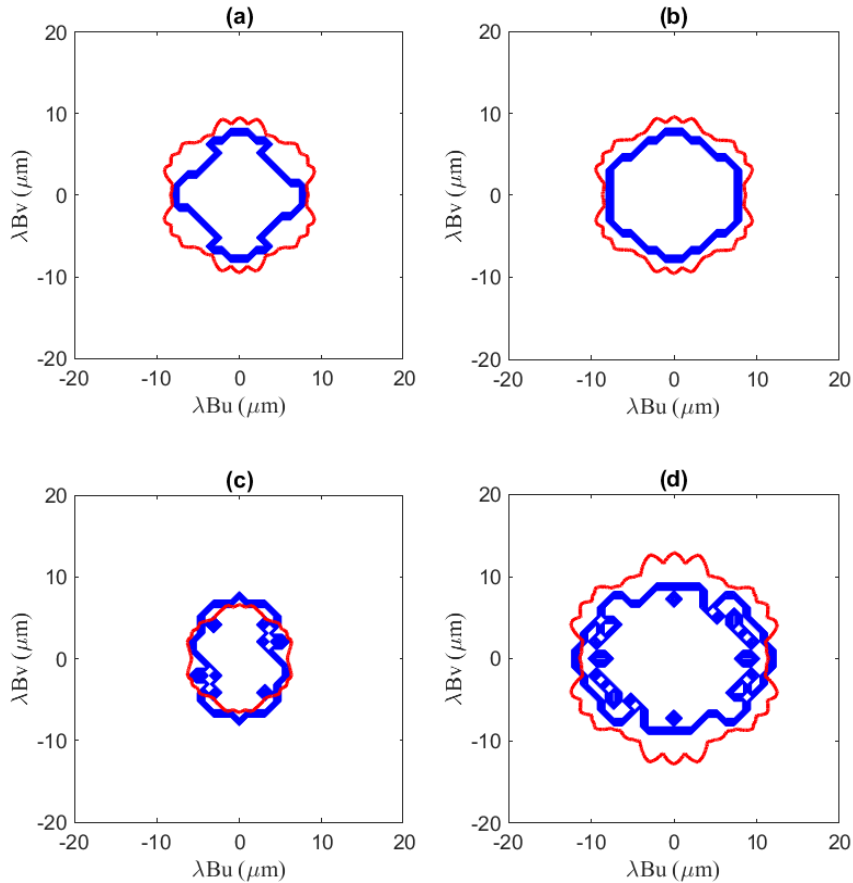

Figure 20: Contours of the 2D-autocorrelation of the particle's shapes (in red) and contours of the 2DFT of the interferometric patterns (in blue) for simple star (a), stellar crystal (b), sectored plate (c) and stellar dendrite (d). $\theta_m = 135°, \phi_m = 0°$, and $\psi = -22.5°$ in all cases.

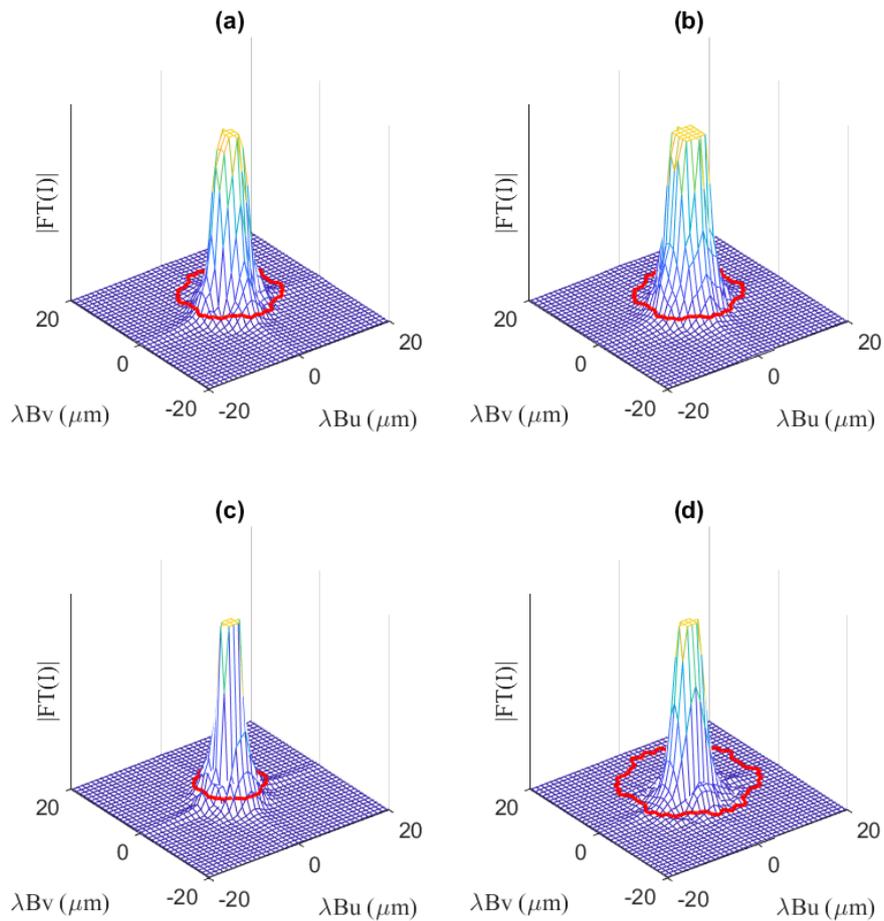

Figure 21: 2DFT of the windowed interferometric patterns and the red contours of the 2D-autocorrelation of the particle's shape for simple star (a), stellar crystal (b), sectored plate (c) and stellar dendrite (d). $\theta_m = 135°$, $\phi_m = 0°$, and $\psi = -22.5°$ in all cases.

## **Conclusion**

Interferometric Particle Imaging is a powerful technique to measure the size and determine the shape of aerosol particles, including ice particles in the atmosphere. Its field of view can exceed tens of square centimeters and it is a single-shot technique; however, its applicability to ice crystal sizes of a few micrometers was questionable. Combining precise ice crystal shape modelling and rigorous light scattering calculations, we have shown that the 2-dimensional Fourier transform of the interferometric pattern remains deeply connected to the 2-dimensional autocorrelation of the shape of the particle. In most forward-scattering imaging cases, this property offers a direct method for sizing and morphological analysis. The only remaining problem is the specular reflections in a backward imaging configuration, which may induce underestimation of the particle size.

For particle sizes of a few micrometers, the detector must be sufficiently large to observe several speckles. In this configuration, a single image contains information from a panel of viewing angles, which must be accounted for in the shape analysis, especially when high-aspect-ratio particles are measured. Moreover, this property may be used to develop a tomographic-like reconstruction from such extended interferometric image, leading potentially to volume estimation. Finally, rigorous DDA simulations offer the possibility to construct large databases of interferometric images to develop future morphological analyses based on deep learning.

**Acknowledgements**

The presented DDA computations were performed using computing resources of CRIANN (Normandy, France). MYu acknowledges support of the Normandy Region (project RADDAERO).

**<u>Appendix A</u>**

Figure A1 shows the reassembly of the distortion-corrected sub-patterns corresponding to the initial magenta, yellow, red and green boxes of figure 7. All four sub-patterns have been recalculated separately with DDA, assuming normal orientation of the sub-detector.

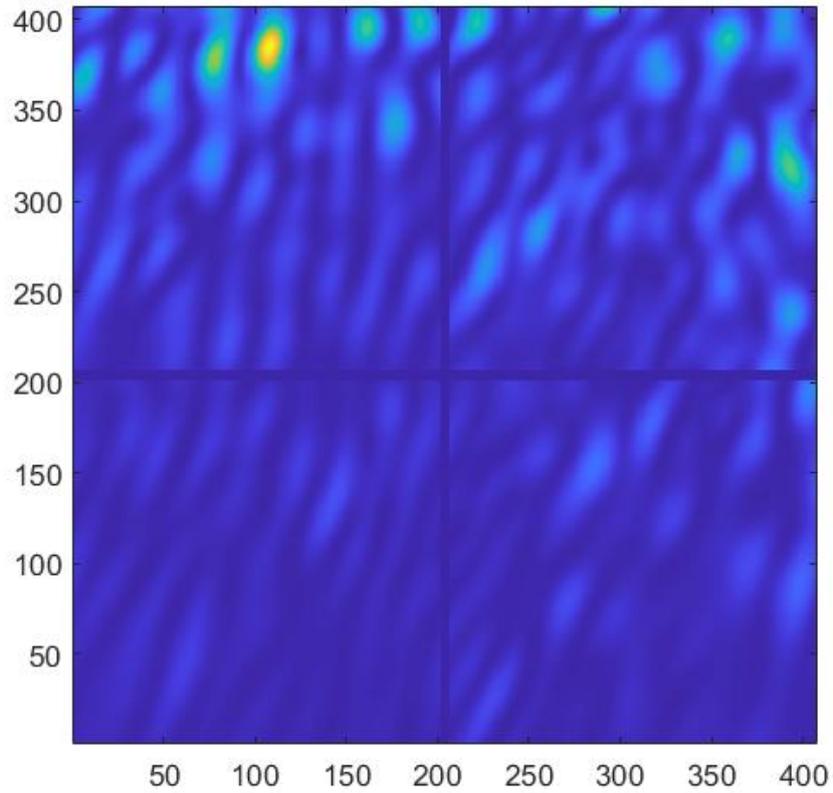

Figure A1: Reassembly of the distortion-corrected sub-patterns corresponding to the initial magenta, yellow, red and green boxes of figure 7. The orientation of the particle is given by $\psi = -22.5°$. The position of the sensor center is given by the pair $(\theta_m = 55°, \phi_m = 90°)$. Axes are in camera's pixels.

**Appendix B**

The different figures of this appendix show the results obtained for particles (simple star, stellar crystal, sectored plate, stellar dendrite) twice larger than those presented in the main text (figures 13-16). Figures B1-B4 correspond to three scattering directions in the forward hemisphere.

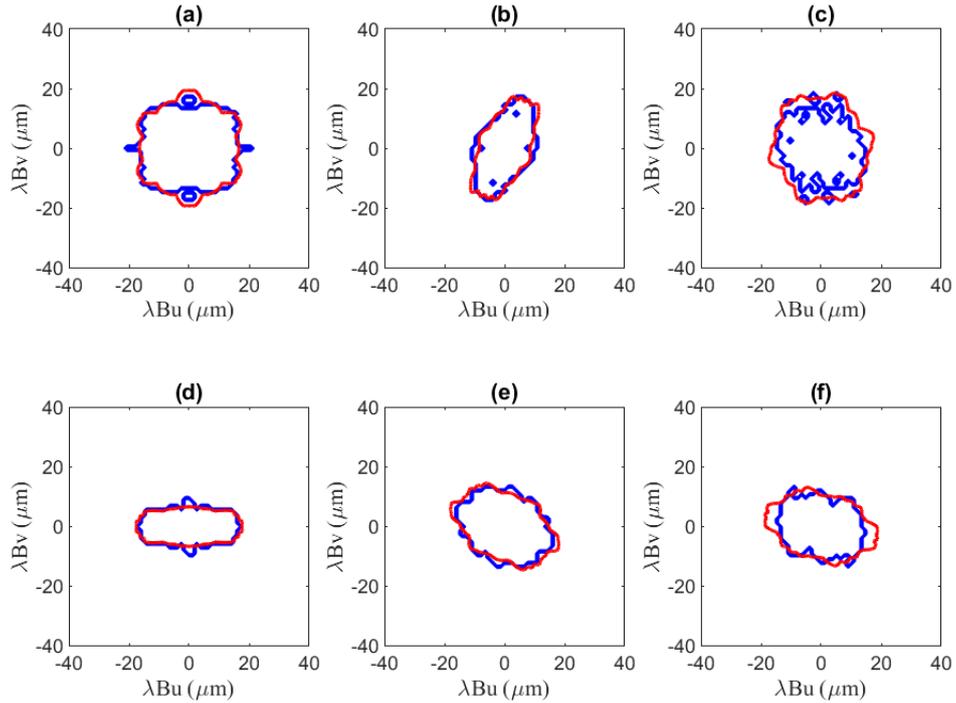

Figure B-1: Contour of the 2D-autocorrelation of the particle's shape (in red) and contour of the 2DFT of the interferometric pattern (in blue) for a simple star ice particle. $\theta_m = 55°, \phi_m = 0°$ in (a) and (d); $\theta_m = 55°, \phi_m = 90°$ in (b) and (e); $\theta_m = 40°, \phi_m = 35°$ in (c) and (f). The particle's orientation is given by $\psi = 55°$ in (a-c) and $\psi = -22.5°$ (d-f).

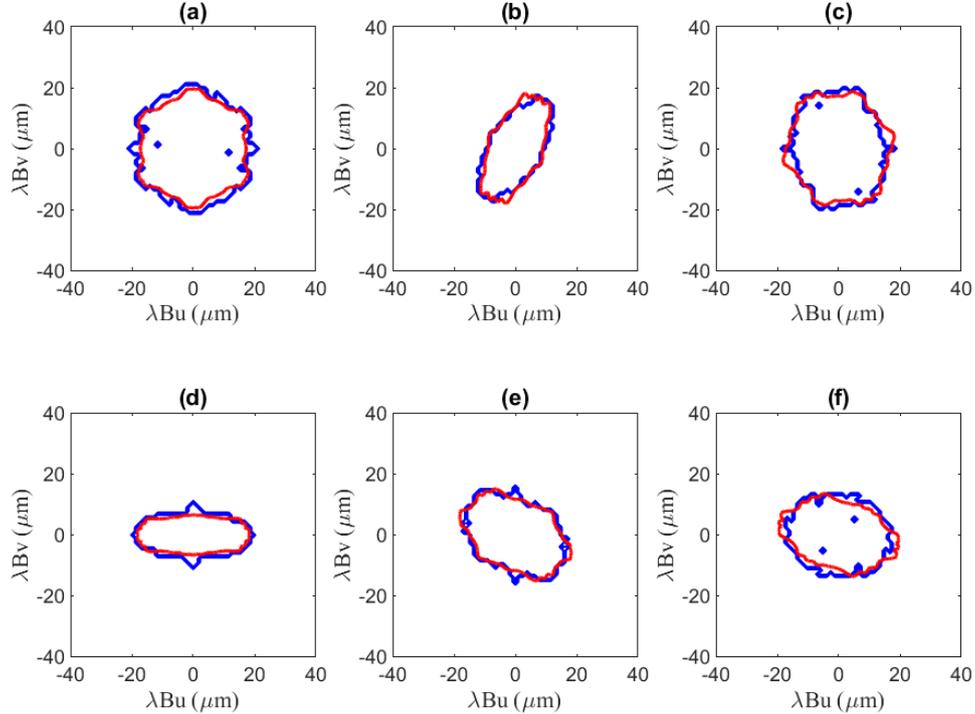

Figure B-2: Contour of the 2D-autocorrelation of the particle's shape (in red) and contour of the 2DFT of the interferometric pattern (in blue) for a stellar ice particle. $\theta_m = 55°, \phi_m = 0°$ in (a) and (d); $\theta_m = 55°, \phi_m = 90°$ in (b) and (e); $\theta_m = 40°, \phi_m = 35°$ in (c) and (f). The particle's orientation is given by $\psi = 55°$ in (a-c) and $\psi = -22.5°$ (d-f).

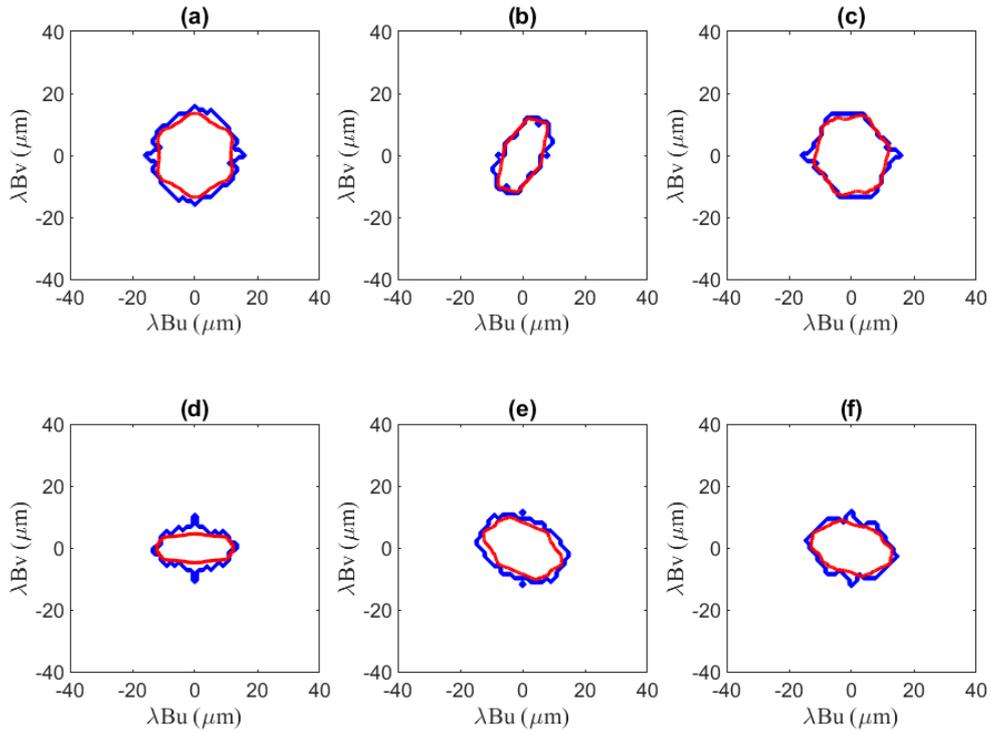

Figure B-3: Contour of the 2D-autocorrelation of the particle's shape (in red) and contour of the 2DFT of the interferometric pattern (in blue) for a sectored plate. $\theta_m = 55°, \phi_m = 0°$ in (a) and (d); $\theta_m = 55°, \phi_m = 90°$ in (b) and (e); $\theta_m = 40°, \phi_m = 35°$ in (c) and (f). The particle's orientation is given by $\psi = 55°$ in (a-c) and $\psi = -22.5°$ (d-f).

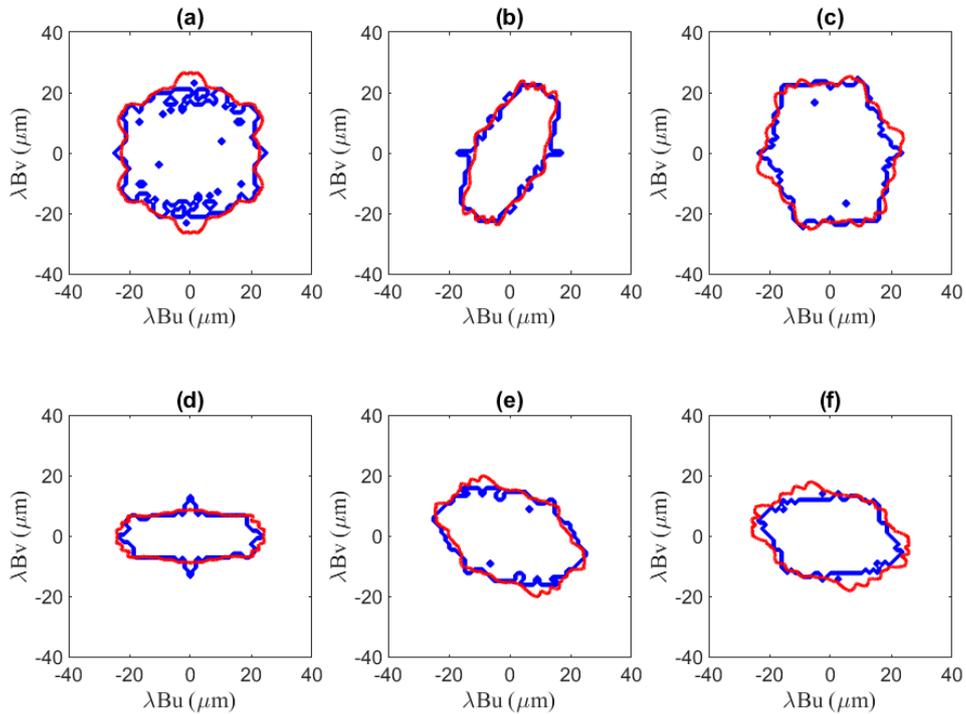

Figure B-4: Contour of the 2D-autocorrelation of the particle's shape (in red) and contour of the 2DFT of the interferometric pattern (in blue) for a stellar dendrite. $\theta_m = 55°, \phi_m = 0°$ in (a) and (d); $\theta_m = 55°, \phi_m = 90°$ in (b) and (e); $\theta_m = 40°, \phi_m = 35°$ in (c) and (f). The particle's orientation is given by $\psi = 55°$ in (a-c) and $\psi = -22.5°$ (d-f).

**Appendix C**

The different figures of this appendix show the results obtained for particles (simple star, stellar crystal, sectored plate, stellar dendrite) in a backward scattering configuration: $\theta_m = 135°, \phi_m = 0°$, and a particle orientation given by $\psi = 55°$. The largest dimensions of particles are around $10 \mu m$ in figure C-1, and twice larger in figure C-2. Particles are almost viewed from the edge in this configuration.

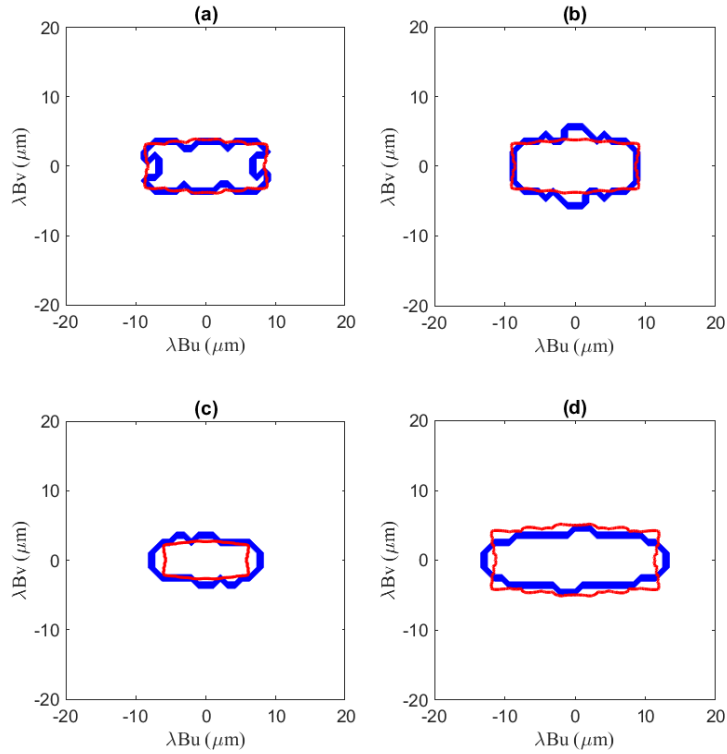

Figure C-1: Contours of the 2D-autocorrelation of the particle's shapes (in red) and contours of the 2DFT of the interferometric patterns (in blue) for simple star (a), stellar crystal (b), sectored plate (c) and stellar dendrite (d). $\theta_m = 135°, \phi_m = 0°$, and $\psi = 55°$ in all cases.

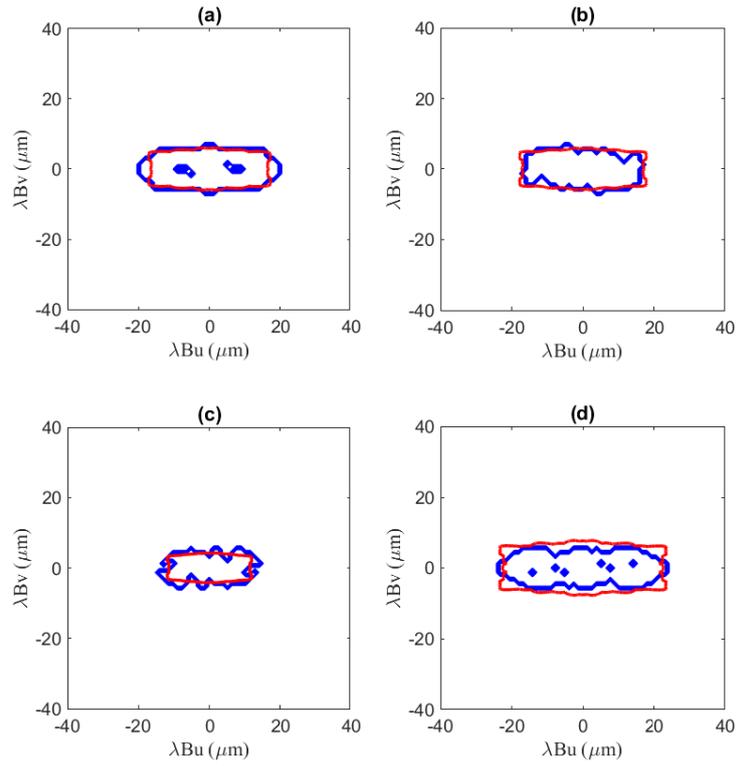

Figure C-2: Contours of the 2D-autocorrelation of the particle's shapes (in red) and contours of the 2DFT of the interferometric patterns (in blue) for simple star (a), stellar crystal (b), sectored plate (c) and stellar dendrite (d). $\theta_m = 135°, \phi_m = 0°$, and $\psi = 55°$ in all cases.